\documentclass{aip-cp}

\usepackage{rotating}
\usepackage{graphicx}

\begin{document}

\title{On the Parallelization of Stellar Evolution Codes}

\author[aff1,aff2]{David Martin}
\author[aff1,aff2]{Jordi Jos\'e}
\author[aff3]{Richard Longland}
\affil[aff1]{Dept. Physics, Universitat Polit\'ecnica de Catalunya, Av. Eduard Maristany 16, 08019 Barcelona.}
\affil[aff2]{Institut d'Estudis Espacials de Catalunya, C. Gran Capit\`a, 2-4, 08034 Barcelona.}
\affil[aff3]{Dept. Physics and Astronomy, North Carolina State University, 421 Riddick Hall, Raleigh, NC 27695-8202.}

\maketitle

\begin{abstract}
Multidimensional nucleosynthesis studies
with hundreds of nuclei linked through thousands of nuclear processes are still computationally prohibitive.
To date,  most nucleosynthesis studies rely either on hydrostatic/hydrodynamic
simulations in spherical symmetry, or on post-processing simulations using temperature and
density versus time profiles directly linked to huge
nuclear reaction networks.

Parallel computing has been regarded as the main permitting factor of computationally
intensive simulations. This paper explores the different pros and cons in the
parallelization of stellar codes, providing recommendations on when
and how parallelization may help in improving the performance of a code for astrophysical applications.

We report on different parallelization strategies succesfully applied to the
spherically symmetric, Lagrangian, implicit hydrodynamic code {\tt SHIVA},
extensively used in the modeling of classical novae and type I X-ray bursts.

When only matrix build-up and inversion processes in the nucleosynthesis subroutines are parallelized (a suitable approach for post-processing calculations),
the huge amount of time spent on communications between cores, together with the small problem size (limited by the number of isotopes of the nuclear network),
 result in a much worse performance of the parallel application compared to the 1-core, sequential version of the code.  Parallelization of the matrix build-up and inversion processes
in the nucleosynthesis subroutines is not recommended unless the number of isotopes adopted largely exceeds 10,000.

In sharp contrast, speed-up factors of 26 and 35 have been obtained with a parallelized version of {\tt SHIVA}, in a 200-shell simulation of a type I X-ray burst carried out with
 two nuclear reaction networks:
a reduced one, consisting of 324 isotopes and 1392 reactions, and a more extended network with 606 nuclides and 3551 nuclear interactions.
Maximum speed-ups of $\sim$41 (324-isotope network) and $\sim$85 (606-isotope network), are also predicted for 200 cores, stressing that
the number of shells of the computational domain constitutes an effective upper limit for the maximum number of cores that could be used
in a parallel application.
\end{abstract}

\section{Introduction}
\label{sec:intro}

Computational astrophysics has revolutionized our knowledge of the 
physics of stars.
Simultaneously to the progress achieved in observational astrophysics 
(through high-resolution spectroscopy and photometry, sometimes including multiwavelength observations with
 space-borne and ground-based observatories), 
cosmochemistry (isotopic abundance determinations 
in presolar meteoritic grains) and nuclear physics 
(determination of nuclear cross sections at or close to stellar energies), 
computers have provided astrophysicists with the appropriate arena in which 
complex physical processes operating in stars (e.g., rotation, convection and mixing, mass loss...) can be properly modeled 
(see, e.g., Ref. [1]).

Stellar evolution models are becoming increasingly sophisticated and complex. The dawn of 
supercomputing and multi-core machines has allowed to (partially) overcome the limitations 
imposed by the assumption of spherical symmetry. The pay-off, however, is still very expensive. 
Two-, and specially three-dimensional simulations are so computationally demanding that other 
simplifications, such as the use of truncated nuclear reaction networks, large enough to account 
for the energetics of the star, must be adopted.
Multidimensional nucleosynthesis studies 
with hundreds of nuclear species linked through thousands of nuclear processes are still prohibitive. 
Accordingly,  most of our understanding of element synthesis in stars relies either on hydrostatic/hydrodynamic 
simulations in spherical symmetry (1D), or on post-processing simulations using temperature and 
density versus time profiles extracted  from stellar evolution models, and directly linked to huge 
nuclear reaction networks. Even such post-processing calculations can sometimes become computationally 
very intensive: for instance, the sensitivity study of the effect of nuclear uncertainties in
X-ray bursts nucleosynthesis performed by Parikh et al. [2], requiring 50,000 post-processing  
calculations, with a network containing 600 species (from H to $^{113}$Xe),
 and more than 3500 nuclear reactions, took about 9 CPU months in a single-core computer.
 
In the 1D codes used in the modeling of a wide range of astrophysical scenarios, such as  classical novae,
X-ray bursts, supernovae, or asymptotic giant branch (AGB) stars
 (e.g., {\tt FRANEC} [3,4], {\tt MESA} [5,6], 
{\tt SHIVA} [7,8]),
 stars are divided into $\sim$ 100s - 1000s of concentric shells.
They also incorporate a similar number of nuclear processes, which link hundreds of nuclear
species. The subroutines that handle the suite of different nuclear processes and the associated
nucleosynthesis are often the most time-consuming components of a stellar evolution code (unless very small nuclear 
reaction networks are used). Different strategies have been adopted to reduce the computational cost
of such simulations, therefore improving the performance of a code. One possibility relies on the use 
of more efficient numerical techniques to handle integration of large nuclear networks [9,10]. 
 Another possibility involves parallelization of the stellar code, 
so that the high computational cost can be split and handled by different cores working cooperatively. 

Parallel computing has been regarded as the main permitting factor of more precise, computationally 
intensive simulations. Indeed, most of the existing multidimensional stellar codes have been parallelized. 
Naively, parallelization simply relies on applying several cores to the solution of
a single problem, so that speed-ups are accomplished by executing independent, non-sequentional
portions of the code. In practice, however, parallelization 
comes with a high cost in both engineering 
and programming efforts. And on top of that, it may turn out that parallelization does not pay off altogether, 
for specific applications. Therefore, the main goal of this paper is to explore the advantages
(and disadvantages) associated with the parallelization of stellar codes, outlining recommendations on when 
and how parallelization may help in improving the performance of a code for astrophysical applications. 
 We discuss speed-up factors ranging between 26 and 35 that allow the execution of hydrodynamic simulations coupled 
to large nuclear reaction networks in affordable times. 

The structure of this paper is as follows: 
different strategies in the parallelization of a stellar evolution code 
 (and of the  matrix build-up and inversion processes in the nucleosynthesis subroutines) 
are described in  Sections 2 and 3. 
Special emphasis is devoted to the expected
speed-ups obtained as a function of the size of the nuclear reaction network and the
number of  cores involved in the simulation. 
The performance of the
parallelized version of {\tt SHIVA} code is qualitatively compared with other codes,
with similar or different architectures, in Section 4.
The main results and 
conclusions of this work, together with a list of open issues, are summarized as well in Section 4.  

\section{Parallelization of a Stellar Code with a  Decoupled, Time-Explicit Treatment of the Nucleosynthesis Subroutines}
\label{sec:hydro_parallelization}

\begin{figure}
\includegraphics[scale=0.4]{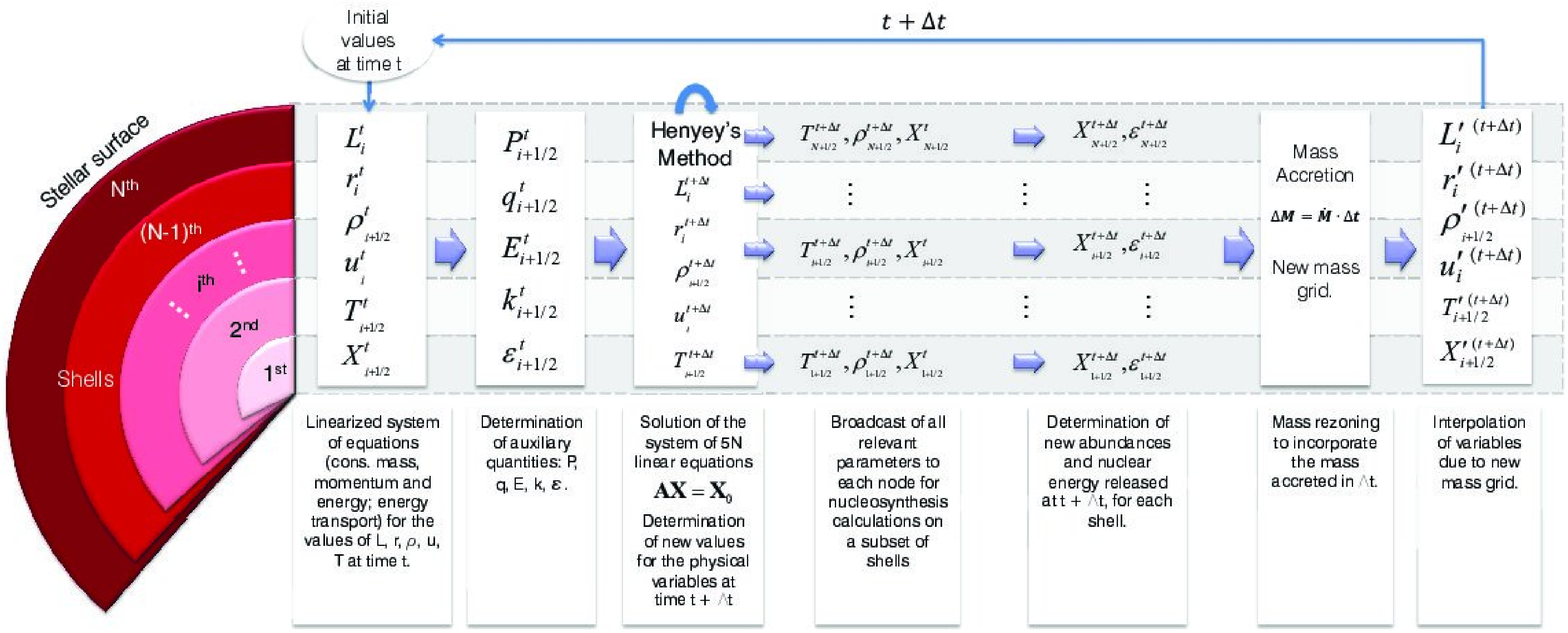}
  \caption{The {\tt SHIVA} code workflow.  The code uses a co-moving (Lagrangian) coordinate system, such that all physical variables 
(luminosity, $L$, velocity, $u$, distance to the stellar center, $r$, density, $\rho$, temperature, $T$, pressure, $P$, 
internal energy, $E$, artificial viscosity, $q$, mass fractions, $X$, opacity, $k$, and energy generation rate, $\epsilon$)
are evaluated in a number of grid points directly attached to the fluid (see Refs. [7,8], for details).}
    \label{fig:Computation_methods}
    \end{figure}

 The different strategies in the parallelization of a stellar evolution code described in this paper rely on the {\it Message Passing Interface} ({\tt MPI}) communication 
protocol, and have been directly applied to {\tt SHIVA},  
a one-dimensional (spherically symmetric), hydrodynamic code, in Lagrangian formulation, built originally to model classical nova outbursts 
(see Refs. [7,8], for details).
The code uses a co-moving (Lagrangian) coordinate system, such that all physical variables (i.e., luminosity, $L$, velocity, $u$, distance to the stellar center, $r$, 
density, $\rho$, and temperature, $T$)
are evaluated in a number of grid points directly attached to the fluid. In essence, this corresponds to a system of $5N$ variables (unknowns), where
$N$ is the  overall number of shells of the computational domain. SHIVA's computational flow is depicted in Fig.~\ref{fig:Computation_methods}.

At each time-step, the set of $5N$ unknowns
is determined from a system of $5N$ linearized equations (i.e., conservation of mass, momentum and energy, the definition of the Lagrangian velocity and an
equation that accounts for energy transport), which is solved by means of an iterative technique---Henyey's method [11].
The basic set of stellar structure equations,  supplemented by a suitable 
 equation of state (EOS, that includes radiation, ions, and electrons with different degrees of
degeneracy), opacities and a nuclear reaction network, constitute the building blocks of any stellar evolution code. 
In {\tt SHIVA}, convection and nuclear energy production
are  decoupled from the set of hydrodynamic equations, and handled by
means of a time-explicit scheme.
 In general, partial differential equations involving 
time derivatives can be discretized in terms of variables evaluated (i.e., known) at the previous
time-step ({\it explicit} schemes) or at the current time-step ({\it implicit} schemes).
Explicit schemes are usually easier to implement than implicit schemes.
However, in explicit schemes the time-step is limited by the 
{\it Courant--Friedrichs--Levy condition} 
 that prevents any 
disturbance traveling at the sonic speed from traversing more than one 
numerical cell, thus leading to unphysical results. 
Implicit schemes allow larger time-steps than explicit schemes, with no
precondition on the time-step, but  they require  
an iterative procedure to solve the system at each step. In {\tt SHIVA}, 
 all compositional changes driven by nuclear processes
or convective transport are evaluated at the end of the iterative 
procedure\footnote{As for the nuclear energy production and nucleosynthesis, 
neutrino losses are also implemented explicitly in the SHIVA code. However, as  they
 do not require intense computation efforts, subroutines handling neutrino losses have not
been parallelized in this work.}, once the temperature, density and the other physical
variables have been determined at each computational shell. In particular, 
{\tt SHIVA} implements a two-step, time-explicit scheme to calculate the new 
chemical composition at each time-step (see Ref. [12]). 
 While such decoupling of the nucleosynthesis subroutines from the hydrodynamic equations
 has a minor effect on the results,
it has a huge impact on the speed-up factors that can be obtained after parallelization (see Section 4, for a more detailed discussion). 

\subsection{Parallelization Strategy}
\label{sec:parallelization_analysis}

The maximum theoretical speed-up accomplished by a parallel application is 
defined as the ratio of the total execution times of the serial application, $T_{\rm S}$,  and the parallel application, $T_{\rm P}$:
\begin{equation}
  \label{eq:speed-up2}
 T_{\rm S}/T_{\rm P} = T_{\rm S}/(T_{\rm in}+T_{\rm pp}/N_{\rm P}+T_{\rm comm}+T_{\rm out}) = 1/((1-p)+p/N_{\rm P}+T_{\rm comm}/T_{\rm S}) 
\end{equation}
where $N_{\rm P}$ is the number of processes participating in the parallel computation, $T_{\rm comm}$ the time devoted to communications and message passing amongst  cores,
$T_{\rm in}$ and $T_{\rm out}$ are the initialization and output times, 
and $p = T_{\rm pp}/T_{\rm S}$ is the so-called {\it parallel content}, or ratio  of the serial execution times of the overall application, $T_{\rm S}$, and
the potentially parallel portion of the code (e.g., a subroutine), $T_{\rm pp}$.
The maximum attainable speed-up\footnote{Note that $T_{\rm S} \equiv T_{\rm in}+T_{\rm pp}+T_{\rm out}$ and $T_{\rm P} \equiv T_{\rm in}+T_{\rm pp}/N_{\rm P}+T_{\rm comm}+T_{\rm out}$.}
is, therefore, determined by the ratio between $T_{\rm comm}$ and $T_{\rm S}$. For $T_{\rm comm} = 0$,  
 Eq.~\ref{eq:speed-up2} results in the well-known Amdahl's law, which  provides an estimate of the
theoretical speed-up as a function of the parallel content and the number of  cores used [13]. 
If the processes need to communicate frequently, the cost of communication will take a 
heavy toll on the total execution time.
 In this situation, speed-ups below unity are even possible 
(i.e., the parallel application will run slower than its sequential counterpart), and therefore, must be avoided.

A first analysis of {\tt SHIVA}'s architecture suggests two main points where parallelization might be exploited: the solution of the  
linearized system of equations for the determination of the physical variables (i.e., Henyey's method), and the multizone calculation of the nuclear energy generation rate
and nucleosynthesis. 
The first one relies on the parallel solution of a system of $5N$ linear equations, where $N$ is the number of shells adopted in the simulation. For a typical
astrophysical application, $N \sim$ 100 - 1,000. However, 
as will be discussed later (see Sect.~\ref{sec:postprocessing_results}), 
such a parallel approach only achieves acceptable performance for $\geq$ 10,000 equations. Very modest speed-up factors are obtained otherwise (i.e., less than a factor of 2),
 which do not justify the effort.
In contrast, the multizone calculation of nuclear energy generation and nucleosynthesis
is computed independently at each shell, and can result in large speed-up factors if parallelized. This is the specific parallelization strategy adopted hereafter, and presented in this Section.
Each  core goes redundantly through almost all processing stages. However, with regard to the nucleosynthesis part, 
each core performs the computation on a non-overlapping subset of shells. After this, each  core broadcasts its (partial) results, and from this stage onward, the simulation proceeds again 
on all  cores redundantly.  In this parallelization strategy adopted, there are only two points of communication: at the beginning of the simulation (where the root process broadcasts all the 
initial information and parameters to the rest of the processes), and repeatedly at each (successful) iteration, after the distributed computation of the nucleosynthesis has been performed. This choice 
maximizes parallel performance by keeping communication points to a minimum or, in other words, by maximizing the computation to communication ratio [14].

In order to obtain equivalent workloads on all cores, the total number of shells of the computational domain must be split up into approximately equally sized groups. 
The shells assigned 
to each  core are consecutive, so that  the different cores compute energy and 
nucleosynthesis for shells $1 \ldots j$, $j+1 \ldots i$, $i+1 \ldots m$, and so on. 
 The last core will have assigned shells $m+1$ to $N$.

\subsection{Performance Prediction}
\label{sec:performance_prediction}

At each iteration, each core broadcasts the new abundances obtained in the computation of their assigned shells. This represents an ALLGATHER communication procedure [15], where 
all processes get the data sent by the other processing cores. The information is thereafter distributed by means of a ring algorithm where, in the first step, each core $i$ sends its contribution 
to  core $i+1$ and receives the contribution from core $i-1$ (with wrap-around). Subsequently, each  core $i$ forwards to  core $i+1$ the data received from core $i-1$ in the previous step 
[16]. The communication time taken by this algorithm is given by [17]:
\begin{equation}
\centering
\label{eq:tcomm}
T_{\rm comm}=(N_{\rm P}-1)\alpha + (N_{\rm P}-1) n\beta/N_{\rm P}
\end{equation}
where $n$ is the total data size received by any  core from all other  cores, $\alpha$ is the latency or start-up time per message (which is independent of the message size), 
and $\beta$ is the transfer time per byte. Actual values for $\alpha$ and $\beta$ obtained in the simulations performed with the {\tt SHIVA} code are given in Sect.~\ref{sec:results_and_discussion}. 
Note that both the latency and the transfer time depend specifically on the speed of the network and of the communications of the computer cluster (or  multi-core computer) 
where the parallel application is being executed. It will also depend on the heterogeneity of the  cores
 (e.g. workstations with different processing power, or different Operating Systems), 
and ultimately on how finely the cluster has been tuned to optimize data transfer and communications. Such quantities are difficult to estimate analytically, and are frequently measured using 
real data and extrapolating communication times from observations [18].

\subsection{Results}
\label{sec:results_and_discussion}

\begin{figure}
\includegraphics{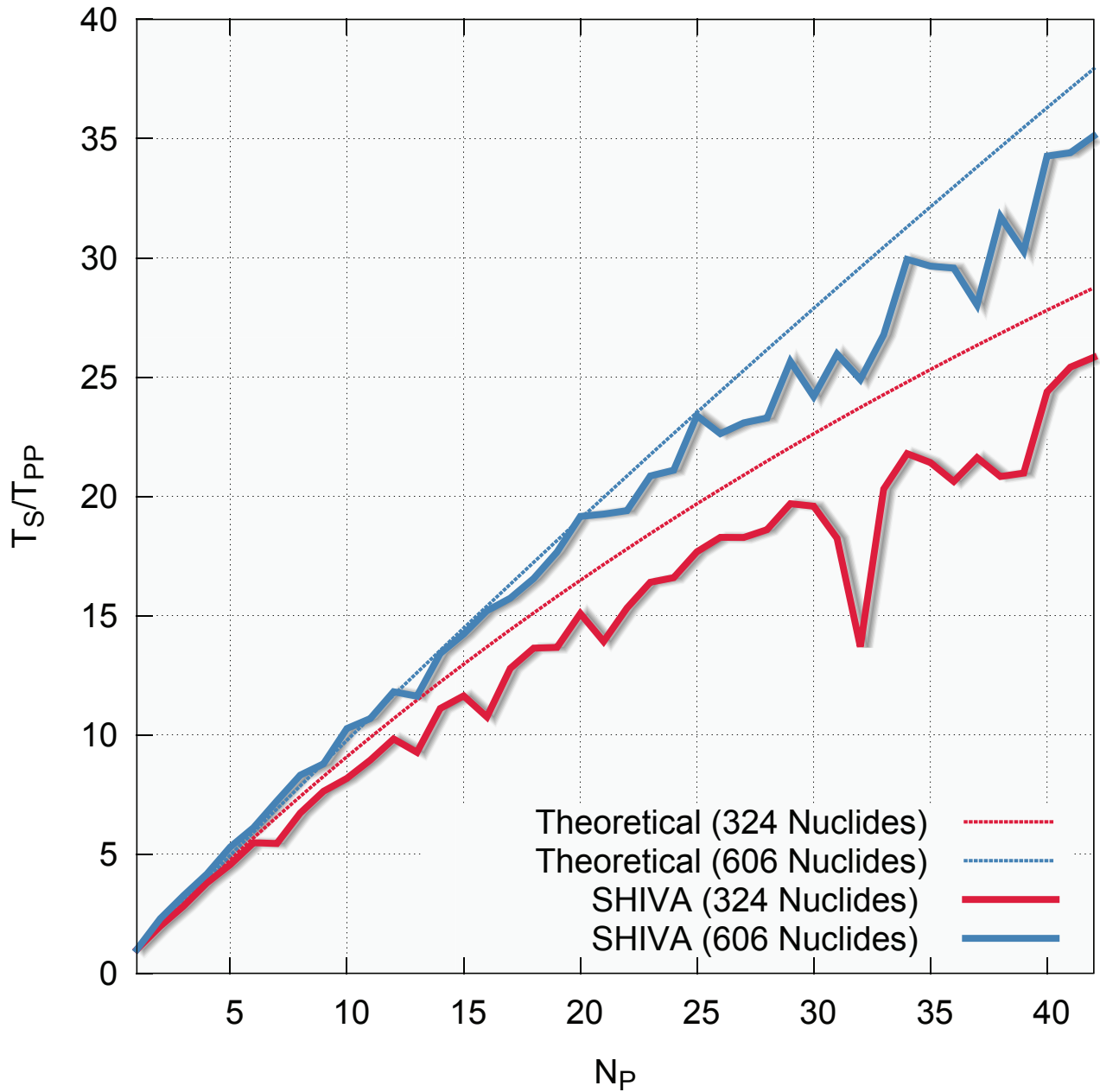}
  \caption{Performance (speed-up factor) of the parallel {\tt SHIVA} code for executions 
with 324 nuclides (Model 1; $p = 0.99127$) and 606 nuclides (Model 2, $p = 0.99738$), 
for $N=200$ shells.
Theoretical speed-ups (thin lines) are compared with real speed-ups obtained 
with the {\tt SHIVA} code (thick lines). 
The theoretical speed-ups correspond to the maximum values expected in the case of a perfect parallelization,
as given by Eq.~\ref{eq:speed-up2}.  }
    \label{fig:shiva_performance1}
    \end{figure}

Fig.~\ref{fig:shiva_performance1} shows the excellent speed-up factors accomplished in a parallel simulation of a type I X-ray burst 
performed with {\tt SHIVA}, with N=200 shells. 
Parallel execution times have been compared with respect to a serial execution time obtained with a single  core. Simulations have been carried out with two different nuclear reaction networks: 
a reduced one, consisting of 324 isotopes and 1392 reactions (hereafter, Model 1), and a more extended network with 606 nuclides and 3551 nuclear interactions (Model 2; see Ref. [19]).
Speed-up factors of 26 and 35 are achieved in Models 1 and 2, respectively, when 42  cores
 are used in parallel to execute the application. 
 Fig.~\ref{fig:shiva_performance1} also highlights the nonlinear scaling of 
the speed-up factor with the number of  cores adopted in the parallel execution.
Both $T_{\rm comm}$ and the overhead time vary with the number of  cores adopted.
This variation depends critically on the type of communication (e.g., all to all,
broadcast, point to point sends and receives, gather, all gather, etc.\footnote{See, e.g.,
https://www.mpi-forum.org/docs/mpi-3.0/mpi30-report.pdf.}),
but at any rate both $T_{\rm comm}$ and the overhead time increase monotonically with the number
of  cores adopted, 
with a much more pronounced dependence of $T_{\rm comm}$ on $N_{\rm P}$ [17].

The results obtained are so good and approach the performance of a perfect parallel application; 
this means that the computation to communication ratio is large enough so that processing work can be distributed in an 
extremely efficient way amongst cores. Accordingly, larger speed-ups are expected if the number of cores 
used in the parallel execution is increased.
Fig.~\ref{fig:shiva_performance1} displays as well the theoretical speed-ups expected for both simulations, 
as given by Eq.~\ref{eq:speed-up2}.
 Such theoretical estimates do not take into account the communication or synchronization 
times, and as a result, the observed performance always falls short compared to the theoretical, ideal speed-up. 
 
 As expected, higher speed-ups are obtained when we increase the problem size by using a 
nuclear reaction network with 606 isotopes and 3551 reactions (e.g., Model 2). 
The speed-up accomplished in this simulation exceeds by approximately 34\% the performance 
of the execution with a reduced nuclear network (i.e., 26 versus 35 speed-up factors, 
respectively). 
This is a direct consequence of increasing the problem size, which is essentially equivalent to 
increasing the amount of parallelizable computation (that is, the nucleosynthesis calculation), 
and therefore the potential parallel content also increases ($p = 0.99127$ for Model 1, whereas $p = 0.99738$ for the simulation with a larger nuclear reaction network, i.e. Model 2). 
This, in turn, improves the curve of the modelled, theoretical speed-up, hence diminishing 
the gap from an ideal speed-up.

\begin{figure}
\includegraphics{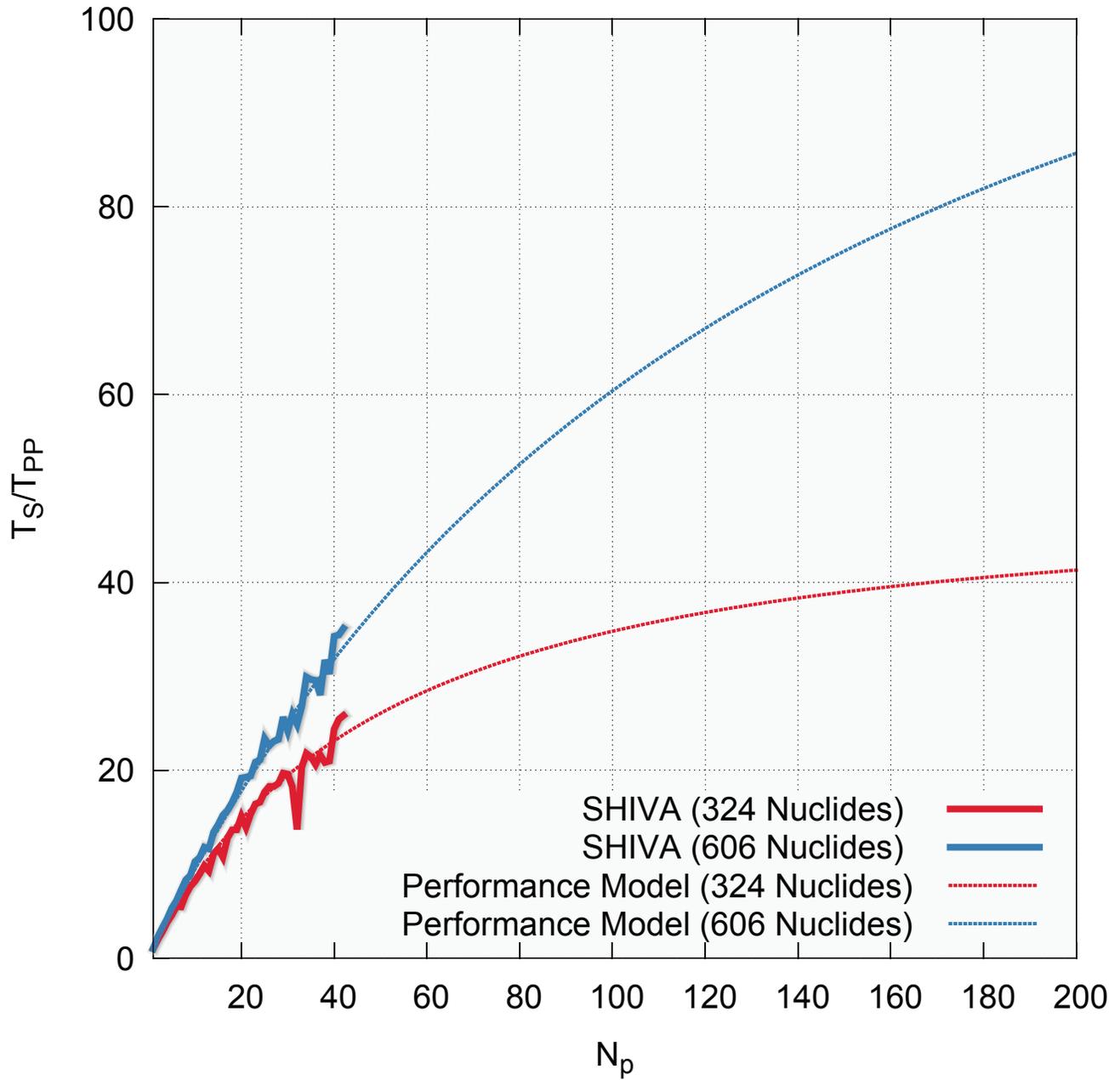}
  \caption{Extrapolated performance model of the parallel {\tt SHIVA} code, with up to 200  cores. A parallel content coefficient of $p = 0.99127$ was used for Model 1 (simulations with
a 324-isotope network), while $p = 0.99738$ was used for Model 2 (606 isotopes).
          The modelled speed-ups correspond to values predicted  for the specific parallelization model discussed in this
          paper and for the {\it Hyperion} cluster, as given by Eq.~\ref{eq:speed-up3}.  }
    \label{fig:shiva_prediction}
\end{figure}

The theoretical performance of the parallelized {\tt SHIVA} code, 
based on Eq.~\ref{eq:speed-up2} and Eq.~\ref{eq:tcomm}, taking into account the communication time between  cores, can be expressed as:
\begin{equation}
  \label{eq:speed-up3}
  {\rm Speed-up} \approx [(1-p)+p/N_{\rm P}+((N_{\rm P}-1)\alpha + (N_{\rm P}-1) n\beta/N_{\rm P})/T_{\rm S}]^{-1} 
\end{equation}
where $n$ and $T_{\rm S}$ are specific of the simulation being executed, and the latency $\alpha$ and the transfer time per byte $\beta$ depend solely on the communications infrastructure. 
Numerical experiments\footnote{All simulations reported in this paper have been executed in the  42-core Hyperion cluster of the Astronomy and Astrophysics Group at UPC.} 
yield $\alpha = 1 \times 10^{-5}$ s and $\beta = 5 \times 10^{-8}$ s. 
At the end of each iteration, all  cores gather the nucleosynthesis results, together with the overall nuclear energy released  and
the values  predicted for the new time-step, $\Delta t$ (e.g., based on the variation of the most abundant isotopes, as in Wagoner's method), 
from all shells. Taking all this into account, the total amount of bytes being transmitted 
works out as:
\begin{eqnarray}
  \label{eq:alphabeta}
200\ {\rm shells} \times (324\ {\rm nuc.} \times 8\ {\rm bytes}/{\rm nuc.} + \nonumber \\
 + 8\ {\rm bytes}/{\rm shell} {\rm (energy)} + 8\ {\rm bytes}/{\rm shell} (\Delta t)) = 521.6 \, \, {\rm kbytes} \nonumber \\
200\ {\rm shells} \times (606\ {\rm nuc.} \times 8\ {\rm bytes}/{\rm nuc.} + \nonumber \\
 + 8\ {\rm bytes}/{\rm shell} {\rm (energy)} + 8\ {\rm bytes}/{\rm shell} (\Delta t)) = 972.8 \, \, {\rm kbytes} \nonumber \\
\end{eqnarray}
for Models 1 and 2, respectively. The expected performance of the parallel {\tt SHIVA} code (Eq.~\ref{eq:speed-up3}) for up to 200  cores is  shown in Fig.~\ref{fig:shiva_prediction}, 
together with the experimental values obtained up to $N_{\rm P}$=42  cores in the {\tt Hyperion cluster}. It is interesting to note that there is still way for improvement.
Indeed, maximum speed-ups of $\sim$41 and $\sim$85 are predicted when using 200  cores, for Models 1 and 2, respectively. 
 The scaling efficiency (i.e., the ratio of actual scaling to ideal scaling) is 21\% for 
Model 1 (speed-up of 41 on 200 cores) and 43\% for Model 2 (speed-up of 85 on 200 cores). 
At this point, it is important to stress
that as a result of the parallelization strategy adopted, the number of shells of the computational domain constitute an effective upper limit for the maximum number of  cores that could be used 
in the parallel application.  Moreover, it is also worth mentioning that the expected performance of the parallel {\tt SHIVA} code, and in general, of any stellar evolution code,
is limited by the number of shells adopted and also by the potentially parallel portion of the code.

It is also important to note that the model of performance presented here is valid for the execution environment discussed, and cannot be extrapolated to other clusters which may have different 
latencies and communication bandwidths. That said, this model can be taken as a reference for the capabilities of a parallelized application, 
and can be used to decide whether  access time at some supercomputing facility, where latencies and transmission bandwidths are highly optimized for parallel executions, must be requested. 
In those platforms, even better speed-up factors must be expected. 

\section{Parallelization of the Nuclear Energy Generation and Nucleosynthesis Subroutines}
\label{sec:postprocessing}

 In this section, we report on the expected speed-ups resulting from parallelization of the matrix
build-up and inversion processes in the nucleosynthesis subroutines, for different sizes of the adopted nuclear reaction networks.
This is a completely different parallelization approach compared to the one described in Section 2. 
In the strategy described for {\tt SHIVA}, the method of solving the system of equations was not modified, but executed in parallel 
on a subset of  non-overlapping shells. Now, it is the build-up and 
inversion of the matrix containing the 
rates of the different nuclear interactions (i.e., the solution of the system of equations) that is being parallelized.
The strategy adopted in this section is of interest for stellar evolution models that rely on reasonably large
nuclear reaction networks, and also for {\it post-processing} nucleosynthesis calculations,  in which 
temperature and density versus time profiles (frequently extracted from stellar models)
are directly coupled to huge nuclear networks. 

\subsection{Numerical Treatment of Nuclear Abundances}
\label{sec:numerical_treatment}

The time-evolution of the chemical composition of a star relies on a set of differential equations that take into account all 
possible creation and destruction channels for the species included in the network. After linearization (e.g., finite-differences),
the overall  system of equations can be written in matrix form as: 
\begin{equation}
\centering
\label{eq:AXXo}
\mathbf{A} \cdot \mathbf{X}=\mathbf{X_0}
\end{equation}
where $\mathbf{X_0}$ is the matrix containing the set of abundances of the previous (or initial) step,
$\mathbf{A}$ is the matrix containing the rates of the different nuclear inteactions, 
and $\mathbf{X}$ is the matrix with the new (unknown) abundances. 

Different methods have been reported to solve Eq.~\ref{eq:AXXo}, such as Wagoner's two-step linearization technique [12], 
Bader-Deuflehard's semi-implicit method [20], or Gear¿s backward differentiation technique [21]. 
The performance of these different integration methods for stellar nucleosynthesis calculations has been been analyzed in a number of
studies (see Refs. [9,10], and references therein). Here, we will explore the gain in performance
driven by parallelization od one particular method: Wagoner's. 
As described in Ref. [22], Wagoner's two-step linearization procedure exploits the special properties
of matrix $\mathbf{A}$, which consists of an upper left square matrix, 
an upper horizontal band, a left vertical band, and a diagonal band. 
The sparse nature of matrix $\mathbf{A}$ results from the fact that the different isotopes, when ordered in terms of increasing atomic number, 
are only linked with close neighbors through nuclear interactions that usually involve light 
particles\footnote{A few exceptions involve reactions such as $^{12}$C + $^{12}$C, $^{16}$O + $^{16}$O,
$^{20}$Ne + $^{20}$Ne, that take place during some stages of the evolution of stars. See Refs. [23,8], for details.}
(e.g., $n$, $p$, $\alpha$).

\subsection{Parallelization Strategy}
\label{sec:ppparallelization}

A typical nucleosynthesis calculation consists of the following main processing steps:
\begin{enumerate}
\item Interpolation (calculation) of reaction rates from tables (analytic fits), for the specific temperature and density of each shell, at a given time.
\item Assembly of matrices $\mathbf{X_0}$ and $\mathbf{A}$. 
\item Solution of Eq.~\ref{eq:AXXo}, for the new abundances of all chemical species at each shell.
\item Convergence check; determination of the new time-step, $\Delta t$.
\item Determination of the overall nuclear energy released at each shell.
\end{enumerate}
Stages 2 and 3 are by far the most time-consuming parts of a simulation (97\% of the execution time in the simulations reported in Sect.~\ref{sec:postprocessing_results}). 
Consequently, the parallelization strategy adopted in this work focused on providing the most efficient partitioning of matrix $\mathbf{A}$, as required by the parallel solution of the system 
of equations performed by the parallel solver.

Reaction-rate determinations are partitioned amongst  cores, such that at each iteration step each core performs the interpolation (calculation) of only those reactions rates that are 
strictly needed for the construction of the local partition of matrix $\mathbf{A}$ (Eq.~\ref{eq:AXXo}). Given a typical nuclear reaction, of the form $i(j,k)l$, 
there are 8 possible  combinations contributing 
to matrix $\mathbf{A}$: $\mathbf{A}(i,i)$, $\mathbf{A}(i,j)$, $\mathbf{A}(j,j)$, $\mathbf{A}(j,i)$, $\mathbf{A}(k,i)$, $\mathbf{A}(k,j)$, $\mathbf{A}(l,i)$, and $\mathbf{A}(l,j)$,
according to the linearization technique described in Ref. [12]. 
The parallel solution of the system of equations is obtained using {\tt MUMPS}\footnote{MUltifrontal Massively Parallel Sparse direct Solver;
see http://mumps.enseeiht.fr/.} [24,25], 
a widely used software for the solution of large sparse systems 
of linear algebraic equations, of the form $\mathbf{A}\mathbf{x}=\mathbf{b}$, on distributed-memory (parallel) computers. 

The right hand side of Eq.~\ref{eq:AXXo} is centralized in the root process. This 
 requires that the complete solution from the previous iteration has to be gathered 
by the root process at some time during the simulation. 
In contrast, the solution of the system of equations is kept distributed, so that after solving the system of equations each of the  cores 
holds a non-overlapping subset of elements of the solution (i.e., a subset of the new abundances). At this point, the solution must be exploited in its distributed form, which requires 
that subsequent processing stages (e.g.,  convergence and accuracy) must be executed independently between  cores.

After solving Eq.~\ref{eq:AXXo}, each  core checks convergence and 
accuracy\footnote{The SHIVA code uses a number of convergence and accuracy criteria to guarantee, for instance, that the new solution satisfies the mass, momentum and energy conservation 
equations.} 
of its part of the solution. 
Finally, the overall nuclear energy released at the specific time-step is obtained by summing the energy generated by all interactions. 
This stage is parallelized by having each  core compute the partial nuclear energy released by a subset of reactions.
The above parallelization strategy requires that the  cores communicate at four specific steps during the simulation:
\begin{enumerate}
\item	During the parallel solution of the system of equations (MUMPS).
\item	Once the system of equations is solved; the distributed solution is shared amongst all 
         cores. 
\item	To check convergence and accuracy of the solution.
\item	To sum up energy contributions from the distributed reactions; every  core computes only the energy released by a subset of reactions.
\end{enumerate}
The above communication requirements are considerably high, as shown by the performance results reported in the following section.

\subsection{Results}
\label{sec:postprocessing_results}

\begin{figure}
\includegraphics{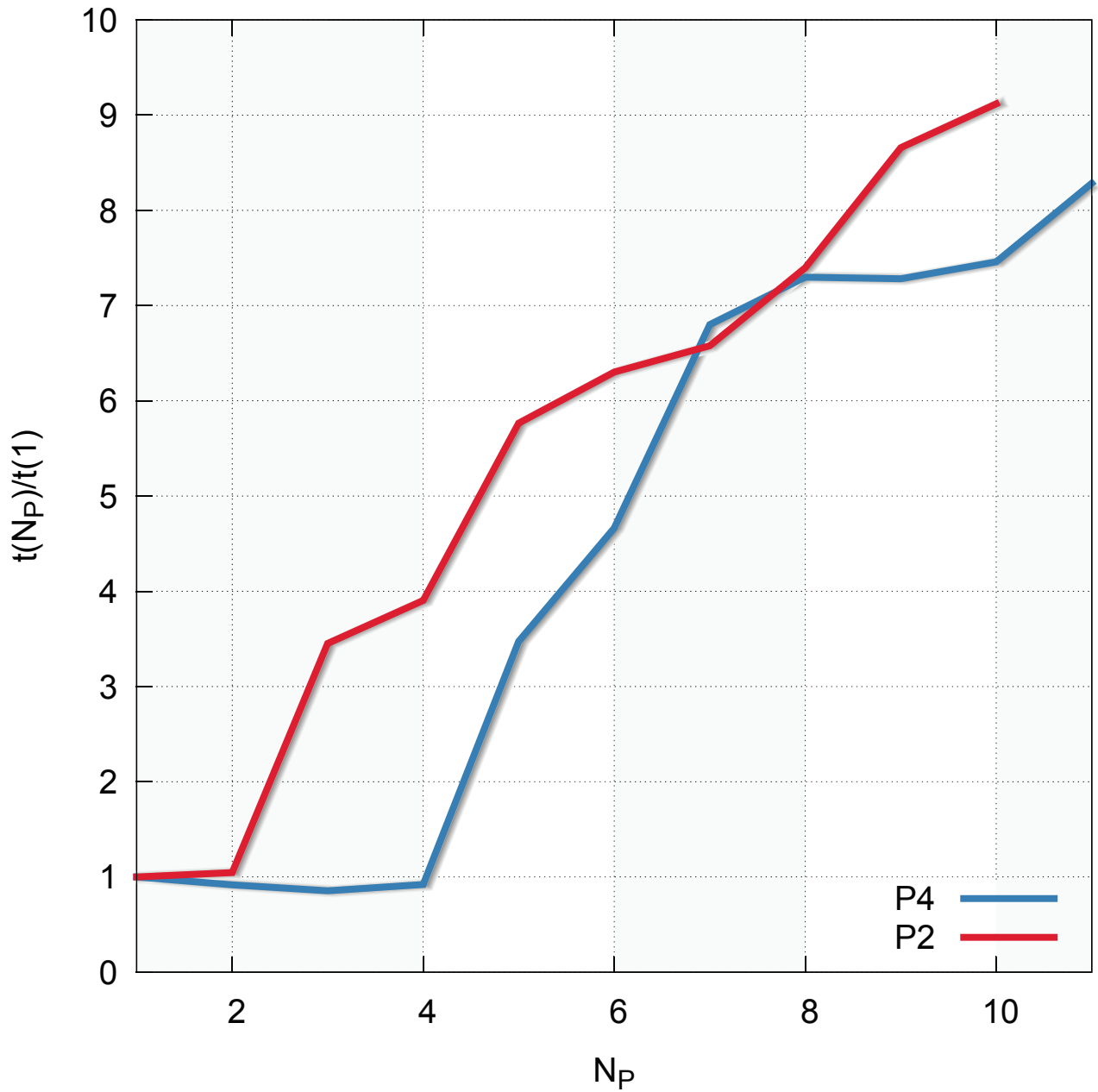}
  \caption{Total execution time as a function of the number of  cores. 
Two different executions are provided, P2 and P4, in which the first two or four cores, respectively, are physically 
located on the same multi-core machine. Execution P2 has been obtained using only dual-core workstations, whereas execution P4 has been run on one quad-core workstation plus 19 
dual-core workstations.}
    \label{fig:executionP4P2}
\end{figure}

The fact that parallelization of the nucleosynthesis subroutines demands much communication between  cores makes the parallel application actually take 
longer to complete than its  1-core counterpart (see Fig.~\ref{fig:executionP4P2}, where the reference value---sequential version---corresponds 
to $N_{\rm P}=1$ and the total execution time is depicted as the ratio between parallel and sequential execution times, $t(N_{\rm P})/t(1)$).

\begin{figure}
\includegraphics[scale=0.6]{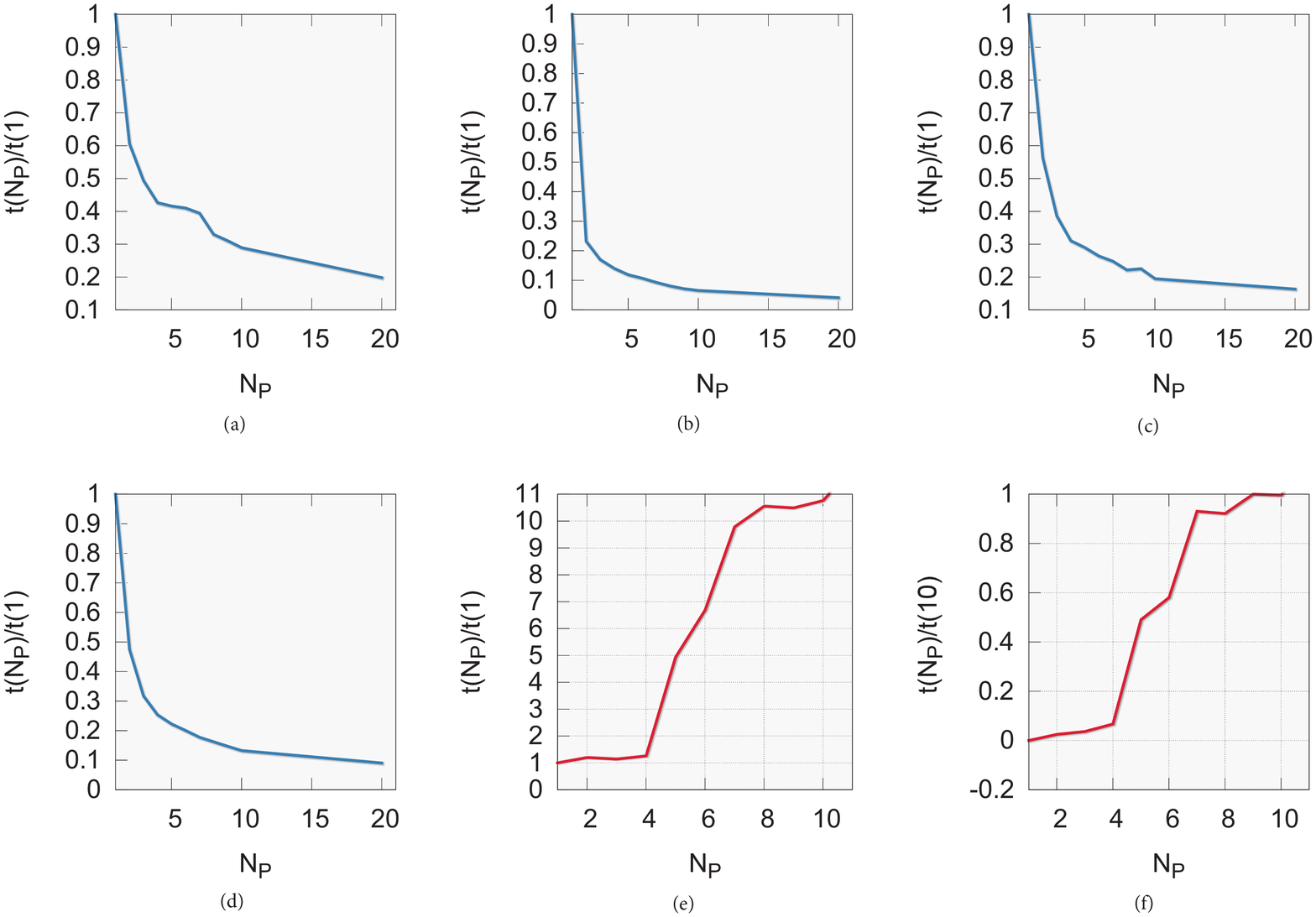}
   \caption{Partial execution times: (a) Rates calculation. (b) Matrix assembly. (c) Convergence check.
   (d) Nuclear energy computation. (e) Matrix build-up and inversion time. (f) Communication time.}
    \label{fig:partial_times}
\end{figure}

The execution time increases when cores physically separated (i.e., on different workstations) participate in the simulation. In sharp contrast, when the parallel application is run 
using cores within the same machine, the execution time is kept at bay with respect to the sequential version, and even small speed-ups are obtained when using a quad core machine, for two, three 
and four  cores.
Fig.~\ref{fig:partial_times} shows the partial execution times spent on the determination of reaction rates (panel a), matrix assembly (b),  convergence check (c), 
and determination of the overall nuclear energy released (d). It is clear that the parallelization strategy adopted for these different
stages is excellent. For instance, the matrix assembly runs almost 5 times faster than the sequential 
version when using 5  cores and almost 7 times faster when using 10  cores. 
The convergence and accuracy check and nuclear energy computation times also yield increases in performance, both running consistently faster in the parallel version than in the sequential application. 
Performance results for the  matrix build-up and inversion time are also shown in Panel e.
It reveals that the solution of the system of equations takes consistently longer if executed in parallel, for any number of  cores used in the computation. Note that for the matrix inversion, 
we do not even get the small improvements when  cores physically located on the same machine are used. Even though the execution time is more or less controlled up to four  cores 
(for a simulation run on a quad-core machine), the performance plummets dramatically with a larger number of  cores. The dramatic loss in performance is therefore provoked 
by the parallel solution of the system of equations. 
The relative time spent on communications is depicted in Panel f. It clearly exhibits the same pattern underlined for the matrix inversion and total execution times.
While the communication time increases slightly from one to four cores, it soars rapidly whenever physically separated cores are incorporated into the parallel execution. 
{\it We conclude that the high communication costs, together with a relatively limited computation time, are responsible for the loss in performance.}

\begin{figure}
\includegraphics{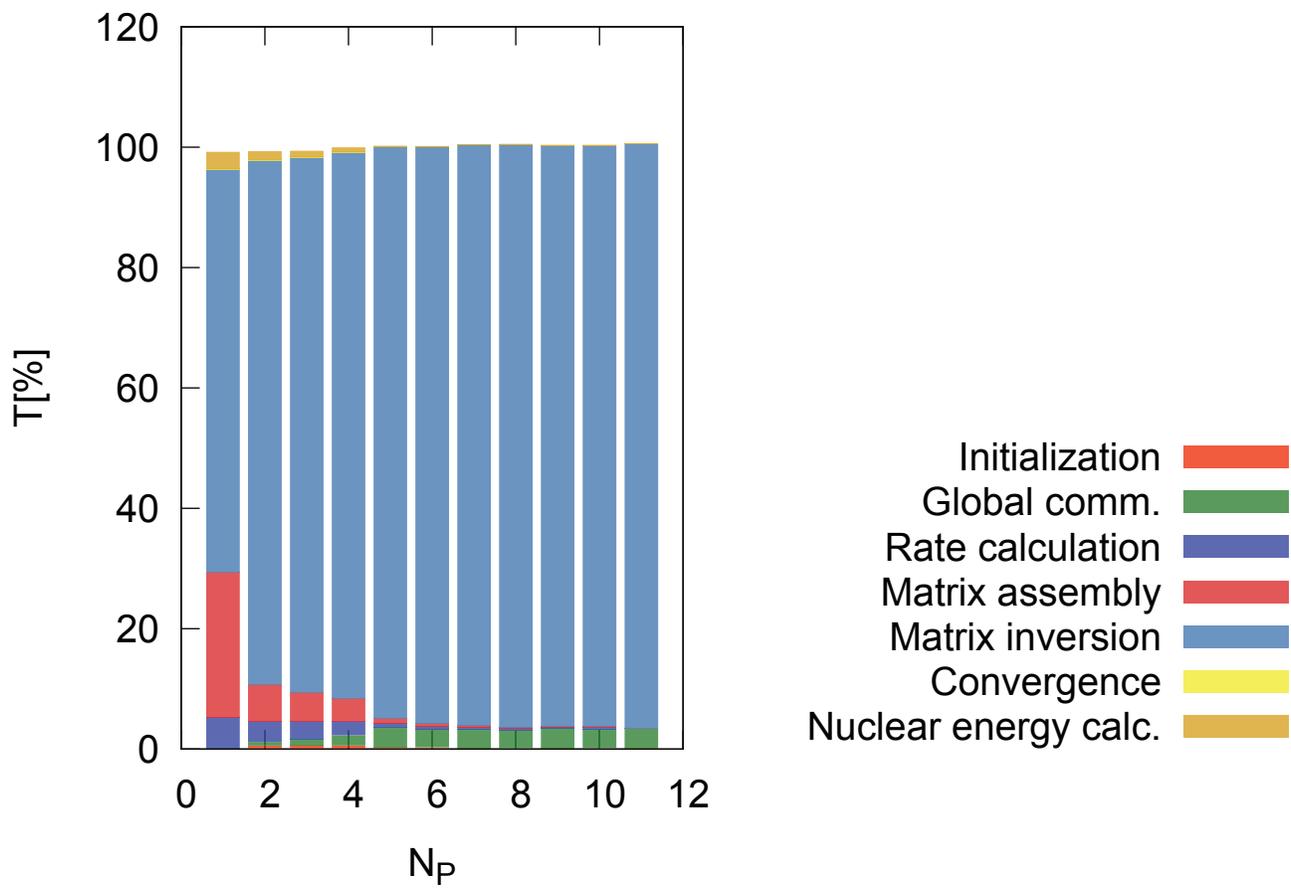}
  \caption{Aggregated simulation time (percentage).}
    \label{fig:percentage_time}
\end{figure}

One final aspect that deserves further discussion is the reason why the gains in performance found in the other stages (see Fig.~\ref{fig:partial_times}) do not make up for the increase 
in communication times. Fig.~\ref{fig:percentage_time} shows the percentage of the total simulation time devoted to initialization, global communications (not including MUMPS internal 
communications during the solution of the system of equations), reaction rate calculations, matrix assembly, convergence check, determination of the nuclear energy released, 
and matrix inversion (i.e., solver). 
The sequential execution spends most of the time inverting the matrix (82\%) and building the system of equations (15\%). The calculation of the overall nuclear energy released accounts only for 
1\% of the total computation time. The relative time spent on the interpolation of reaction rates is just a 0.44\% of the total execution time, whereas only 0.06\% is spent on 
 convergence checks. With an increasing number of  cores participating in the simulation, the time spent on global communications and in the solution of the system of equations gradually tends to 
account for nearly all the computation time. This is the reason why improvements in performance in these 
stages have no major effect on the overall execution time.

Having such a loss in performance associated with the solution of the system of equations, it is compulsory to analyze whether the selection of MUMPS as a solver has been appropriate. 
MUMPS represents one of the few professional and supported public domain implementations of the multifrontal method. Amestoy et al. [24] have shown that the MUMPS solver performance for 
large matrices is excellent. For matrices of order $\geq$100,000, very good speed-ups are accomplished (e.g., between 2.8  and to 3.7, with 4 cores; and between 7.1 and 10.6, with 16 cores). 
Note that speed-ups increase with the matrix size as the computation to communication ratio increases. 
For matrices of the order between 10,000 and 100,000, moderate speed-ups are accomplished with MUMPS (e.g., 2.4 - 3.1, with 4 cores, 
and 7.2 - 8.4, with 16 cores; [26]). Finally, not much data is available for matrices of order $\leq$10,000. 
This is due to the fact that as the problem dimension shrinks, the distributed computation time is also reduced, whilst communication time diminishes much less noticeably. Accordingly,
the resulting speed-ups are dramatically reduced. For instance, Fox [27], in solving a system with 5,535 elements with the MUMPS solver,
  reports speed-ups of 1 (i.e., no speed-up at all) with 4 cores, and a speed-up of 1.8 for 16 cores. 
 It seems clear that the poor performance reported in this work is mostly due to 
the size (order) of the nucleosynthesis matrix, too small to maximize the ratio between computation and communication times.
{\it Accordingly, the efficient parallelization 
 of the matrix build-up and inversion processes in the nucleosynthesis subroutines
is therefore not possible,
unless  $\geq$ 10,000 nuclear interactions are included}. 

\section{Conclusions}
\label{sec:conclusions}

This paper reports on several  parallelization strategies that can be applied to stellar evolution codes, providing recommendations on when 
and how parallelization may help in improving the performance of a code for astrophysical applications. 
Parallelization frequently forces to think about a program in new ways and may virtually require partial or total rewriting of the serial code. 
It is therefore important to understand the potential benefits and risks beforehand, since sometimes 
parallelized codes  may perform even worse than their sequential counterparts.
  
To this end, two different parallelization strategies have been reported in this work.
With regard to the nucleosynthesis part,
efforts have focused on the parallelization of the solution of the system of 
equations (that is, 
 the build-up and inversion of the matrix containing the rates of the different nuclear interactions).
In Wagoner's two-step linearization technique, the integration method for stellar nucleosynthesis calculations discussed in this work, 
the iterative procedure places this application in the worst possible category for parallelization, 
in which all cores have to participate throughout the iteration, exchanging intermediate results on a regular basis. 
The huge amount ot time spent on communications between cores, 
together with the small problem size (limited by the number of isotopes of the nuclear 
network), 
 result in a much worse performance of the parallel application than the 1-core, sequential version of the code.
 This stems from the fact that the communication and message passing times between processes largely outgrow 
the time spent on computation. It is therefore not advisable to parallelize the nucleosynthesis portion of a stellar code (or, by extension, a post-processing code)
unless the number of isotopes adopted largely exceeds 10,000.

With regard to the parallelization of a complete stellar evolution code, efforts have focused on 
the spherically symmetric, Lagrangian, implicit hydrodynamic code {\tt SHIVA} [7,8], 
in the framework of a 200-shell simulation of a typical type I X-ray burst. 
Two different nuclear reaction networks have been considered: 
a reduced one, consisting of 324 isotopes and 1392 reactions; and a more extended network, with 606 nuclides and 3551 nuclear interactions. 
The performance of the parallelized version of {\tt SHIVA} 
turned out to be  excellent: speed-up factors of 26 and 35 have been obtained, 
for the reduced (i.e., Model 1) and extended networks (Model 2), 
respectively, when 42 cores were used. 
These results, however,  did not match the maximum expected values for a perfect parallel application (i.e., the computation to communication ratio was large enough 
so that processing work could be distributed in an extremely efficient way amongst processes). 
To put these results into context, in our execution environment, a parallel simulation using 42 cores took $\sim$ 5.7 hr to compute 200,000 time-steps with a reduced nuclear network 
(c.f., 6.1 days in its sequential version). The computation time increased to $\sim$ 20 hr when the extended network (with 606 nuclides and 3551 nuclear reactions) was used, 
for the same number of time-steps (c.f., 28.6 days in its sequential version). Such excellent results completely justify the time invested in the parallelization of the code. 
Moreover, maximum speed-ups of $\sim 41$ and $\sim 85$ have been predicted by the performance model when using 200 cores, for the reduced and extended nuclear networks, respectively.

A key ingredient in achieving the large speed-up factors reported above is the 
decoupling of the nucleosynthesis subroutines from the set of hydrodynamic/structure 
equations adopted in {\tt SHIVA}. This approach,
while having a minor effect on the expected energetics and chemical composition of
a star, is essential to justify a parallelization effort. In sharp contrast, efforts to 
parallelize {\tt FRANEC} (see Refs. [3,4], and references therein), 
another Henyey-type code in which the nucleosynthesis and structure equations 
are solved simultaneously by means of a time-implicit scheme\footnote{More recent versions
of the FRANEC code, known as FUNS, contain several solver schemes in which 
the equations of nucleosynthesis, mixing and structure can be handled in a 
coupled or decoupled way (O. Straniero, private com.). The extensively used 
 MESA code [5,6] also solves the nucleosynthesis and composition equations 
directly coupled to the structure equations. Note, however, that  MESA contains a
number of explicit modules that can be computed in parallel using  OpenMP [5].}, 
yielded very poor speed-up factors 
(A. Chieffi, private com.). 

In summary, parallelization of a fully coupled, time-implicit code can only result 
in large speed-factors if the most time-consuming parts of the code (e.g., the 
nucleosynthesis subroutines) are decoupled
from the hydro equations, and therefore, can be handled in a time-explicit way.
Most multidimensional, stellar evolution codes available to date
(e.g., {\tt PROMETHEUS} [28]; {\tt FLASH} [29]; {\tt DJEHUTY}
[30,31]; {\tt GADGET2} [32]) are (time) explicit.
While, in general, explicit schemes are easier to implement than implicit schemes, the
real pay-off is the huge speed-up factors achievable when parallelized, 
compared with their 1-core, sequential versions.

\section{ACKNOWLEDGMENTS}
 This article benefited from discussions within
the COST Action ``Chemical Elements as Tracers of the Evolution of the Cosmos" (ChETEC, CA16117).

\section{REFERENCES}

\noindent
[1] P. Bodenheimer, G. P. Laughlin, M. R\'ozyczka, and H. W. Yorke, Numerical Methods in Astrophysics: An
Introduction (CRC/Taylor and Francis, Boca Raton (FL), USA, 2006).

\noindent
[2] A. Parikh, J. Jos\'e, F. Moreno, and C. Iliadis, Astrophys J Suppl 178, 110-136 (2008).

\noindent
[3] M. Limongi and A. Chieffi, Astrophys J 592, 404-433 (2003).

\noindent
[4] A. Chieffi and M. Limongi, Astrophys J 764, p. 21 (36 pp) (2013).

\noindent
[5] B. Paxton, L. Bildsten, A. Dotter, F. Herwig, P. Lesaffre, and F. Timmes, Astrophys J Suppl 192, p. 3 (35
pp) (2011).

\noindent
[6] B. Paxton, M. Cantiello, P. Arras, L. Bildsten, E. F. Brown, A. Dotter, C. Mankovich, M. H. Montgomery,
D. Stello, F. X. Timmes, and R. Townsend, Astrophys J Suppl 208, p. 4 (42 pp) (2013).

\noindent
[7] J. Jos\'e and M. Hernanz, Astrophys J 494, 680-690 (1998).
 
\noindent
[8] J. Jos\'e, Stellar Explosions: Hydrodynamics and Nucleosynthesis (CRC/Taylor and Francis, Boca Raton (FL),
USA, 2016).

\noindent
[9] F. X. Timmes, Astrophys J 124, 241-263 (1999).

\noindent
[10] R. Longland, D. Martin, and J. Jos\'e, Astron Astrophys 563, p. A67 (13 pp) (2014).

\noindent
[11] L. G. Henyey, J. E. Forbes, and N. L. Gould, Astrophys J 139, 306-317 (1964).

\noindent
[12] R. V. Wagoner, Astrophys J Suppl 18, 247-295 (1969).

\noindent
[13] G. M. Amdahl, Validity of the Single Processor Approach to Achieving Large Scale Computing Capabilities, 
  in Proceedings of the April 18-20, 1967, Spring Joint Computer conference (New York (NY), USA:
ACM Publ., 1967), pp. 483-485.

\noindent
[14] P. E. McKenney, Is Parallel Programming Hard, And, If So, What Can You Do About It? (Paper Linux
Technology Center, New York (NY), USA, 2011).

\noindent
[15] R. Graham, MPI: A Message-Passing Interface Standard, v3.0 (University of Tennessee, Tech. Rep.,
Knoxville (TN), USA, 2012).

\noindent
[16] P. Pacheco, Parallel Programming with MPI (Morgan Kaufmann Publ., San Francisco (CA), USA, 1997).

\noindent
[17] R. Thakur and W. Gropp, Improving the Performance of MPI Collective Communication on Switched
Networks, in Recent Advances in Parallel Virtual Machine and Message Passing Interface, edited by D. L.
J Dongarra and S. Orlando (Berlin, Germany:Springer, 2003), pp. 257-276.

\noindent
[18] I. Foster, Designing and Building Parallel Programs (Addison-Wesley, Boston (MA), USA, 1995).

\noindent
[19] J. Jos\'e, F. Moreno, A. Parikh, and C. Iliadis, Astrophys J Suppl 189, 204-239 (2010).

\noindent
[20] G. Bader and P. Deuflhard, Numerische Mathematik 41, 373-398 (1983).

\noindent
[21] C. W. Gear, Commun ACM 14, 176-179 (1971).

\noindent
[22] N. Prantzos, M. Arnould, and J. P. Arcoragi, Astrophys J 315, 209-228 (1987).

\noindent
[23] C. Iliadis, Nuclear Physics of Stars (2nd Ed.) (Wiley-VCH Verlag, Weinheim, Germany, 2015).

\noindent
[24] P. R. Amestoy, I. S. Duff, J. Y. L'Excellent, and X. S. Li, Performance and Tuning of Two Distributed
Memory Sparse Solvers, in Proceedings of the 10th SIAM Conference on Parallel Processing for Scientific
Computing, edited by J. Meza and C. Koelbel (Portsmouth (VA), USA: Society for Industrial \& Applied
Mathematics, 2001a).

\noindent
[25] P. Amestoy, A. Guermouche, J. Y. L'Excellent, and S. Pralet, Hybrid Scheduling for the Parallel Solution of
Linear Systems (CERFACS, Tech. Rep., Toulouse, France, 2004).

\noindent
[26] P. R. Amestoy, I. S. Duff, J. Y. L'Excellent, and J. Koster, SIAM J Matrix Anal Appl 23, 15-41 (2001b).

\noindent
[27] J. Fox, Fully-Kinetic PIC Simulations for Hall-Effect Thrusters, Ph.D. thesis, Massachusetts Institute of
Technology 2007.

\noindent
[28] B. Fryxell, E. Müller, and W. D. Arnett, Hydrodynamics and Nuclear Burning (Max-Planck Inst. for Astrophysics, Rep. 449, Garching, Germany, 1989).

\noindent
[29] B. Fryxell, K. Olson, P. Ricker, F. X. Timmes, M. Zingale, D. Q. Lamb, P. MacNeice, R. Rosner, J. W. Truran,
and H. Tufo, Astrophys J Suppl 131, 273-334 (2000).

\noindent
[30] D. S. P. Dearborn, J. R. Wilson, and G. J. Mathews, Astrophys J 630, 309-320 (2005).

\noindent
[31] D. S. P. Dearborn, J. C. Lattanzio, and P. P. Eggleton, Astrophys J 639, 405-415 (2006).

\noindent
[32] V. Springel, Monthly Not Royal Astron Soc 364, 1105-1134 (2005).

\newpage 

\end{document}